\documentclass[(superscriptaddress,nofootinbib,12pt]{revtex4}
\pdfoutput=1
\usepackage{amssymb,amsmath,amsthm,graphicx}
\usepackage{latexsym}
\usepackage{bm}
\usepackage[]{hyperref}
\usepackage{cleveref}

\usepackage{nccmath}	%even more additional math symbols%

\newcommand{\NPl}{\bm{\ell}}					%bold letters for NP tetrad%
\newcommand{\NPn}{\bm{n}}
\newcommand{\NPm}{\bm{m}}
\newcommand{\NPbm}{\bm{\overline{m}}}
\def\be{\begin{equation}}
\def\ee{\end{equation}}
\def\ba{\begin{eqnarray}}
\def\ea{\end{eqnarray}}

\begin{document}

\title{Strong cosmic censorship in de Sitter space}

\author{Oscar~J.~C.~Dias}
\email{ojcd1r13@soton.ac.uk}
\affiliation{STAG research centre and Mathematical Sciences, University of Southampton, UK}

\author{Felicity~C.~Eperon}
\email{fce21@cam.ac.uk}
\affiliation{Department of Applied Mathematics and Theoretical Physics, University of Cambridge, Wilberforce Road, Cambridge CB3 0WA, UK} 

\author{Harvey~S.~Reall}
\email{hsr1000@cam.ac.uk}
\affiliation{Department of Applied Mathematics and Theoretical Physics, University of Cambridge, Wilberforce Road, Cambridge CB3 0WA, UK} 

\author{Jorge~E.~Santos}
\email{jss55@cam.ac.uk}
\affiliation{Department of Applied Mathematics and Theoretical Physics, University of Cambridge, Wilberforce Road, Cambridge CB3 0WA, UK} 

\begin{abstract}
Recent work indicates that the strong cosmic censorship hypothesis is violated by nearly extremal Reissner-Nordstr\"om-de Sitter black holes. It was argued that perturbations of such a black hole decay sufficiently rapidly that the perturbed spacetime can be extended across the Cauchy horizon as a weak solution of the equations of motion. In this paper we consider the case of Kerr-de Sitter black holes. We find that, for any non-extremal value of the black hole parameters, there are quasinormal modes which decay sufficiently slowly to ensure that strong cosmic censorship is respected. Our analysis covers both scalar field and linearized gravitational perturbations. 
\end{abstract}

\maketitle
%%%%%%%%%%%%%%%%%%%%%%%%%%%%%%%%%%%%%%%%%%%%%%%%%%%%%%%%%%%%
%%%%%%%%%%%%%%%%%%%%%%%%%%%%%%%%%%%%%%%%%%%%%%%%%%%%%%%%%%%%
\section{Introduction}
%%%%%%%%%%%%%%%%%%%%%%%%%%%%%%%%%%%%%%%%%%%%%%%%%%%%%%%%%%%%
%%%%%%%%%%%%%%%%%%%%%%%%%%%%%%%%%%%%%%%%%%%%%%%%%%%%%%%%%%%%
The strong cosmic censorship conjecture \cite{penrose} asserts that, for generic asymptotically flat initial data for Einstein's equation, the maximal Cauchy development is inextendible, \emph{i.e.}, Cauchy horizons do not form. It is well-known that the presence of a Cauchy horizon inside an asymptotically flat charged or rotating black hole does not constitute a violation of strong cosmic censorship because of an infinite blue shift at the Cauchy horizon, which renders it unstable and therefore non-generic \cite{Simpson:1973ua,Poisson:1990eh,Dafermos:2003wr,Luk:2017jxq}. Some time ago, it was observed that the mechanism behind this instability is weaker when the cosmological constant $\Lambda$ is positive \cite{Mellor:1989ac}. This is because there is a redshift of late time perturbations entering the black hole, arising from the fact that these perturbations have to climb out of the gravitational potential well associated with the cosmological horizon. 

Early calculations (reviewed in Ref. \cite{Chambers:1997ef}) indicated that, for charged or rotating black holes sufficiently close to extremality, a violation of strong cosmic censorship would indeed be possible with positive $\Lambda$. However, subsequent work argued that the decay of scalar field perturbations outside the black hole was still sufficiently slow to ensure that the gradient of the scalar field would diverge at the Cauchy horizon, with backreaction then causing a curvature divergence, and so strong cosmic censorship would be respected \cite{Brady:1998au}. 

Recent interest in this topic has been stimulated by the recognition that a divergence in curvature does not necessarily imply that spacetime cannot be extended beyond the Cauchy horizon. For $\Lambda=0$, it is always possible to extend the perturbed solution so that the metric (and scalar field) are continuous across the Cauchy horizon \cite{Dafermos:2003wr}. It has also been known for a long time that the divergence in curvature can be sufficiently weak that extended objects may be able to cross the Cauchy horizon without being destroyed \cite{Ori:1991zz}. On the other hand, a divergence in curvature seems problematic because if the metric is not $C^2$ then how could the Einstein equation be satisfied at the Cauchy horizon?

The new interest in this topic stems from the fact that one can still make sense of the Einstein equation even if the metric is not $C^2$. A metric with lower regularity may still constitute a {\it weak} solution of the Einstein equation. The notion of of a weak solution is not just of mathematical interest: physical phenomena, such as shocks in a fluid, are described by weak solutions of equations of motion. For the Einstein equation, the appropriate regularity of weak solutions was determined by Christodoulou \cite{Christodoulou:2008nj}: in some chart the metric should have locally square integrable Christoffel symbols. Therefore the modern statement of the strong cosmic censorship conjecture is that, although it may be possible to extend the metric continuously across the Cauchy horizon, generically it should not be possible to do so with locally square integrable Christoffel symbols \cite{Christodoulou:2008nj}. 

For $\Lambda=0$, it seems very likely that this conjecture is true (see Ref. \cite{Dafermos:2012np} for a detailed discussion). However, with $\Lambda>0$ it was observed in  \cite{Dafermos:2012np} that calculations similar to those of \cite{Brady:1998au} suggest that Christodoulou's version of the strong cosmic censorship conjecture may be false for near-extremal Reissner-Nordstr\"om-de Sitter and Kerr-de Sitter black holes. Very recently, Ref. \cite{Cardoso:2017soq} has presented compelling evidence that this is indeed the case for near-extremal Reissner-Nordstrom de Sitter. The argument is based on recent mathematical developments in the study of black holes with positive $\Lambda$, as we will now explain. 

\begin{figure}
	\centering
	\includegraphics[width=0.45\textwidth]{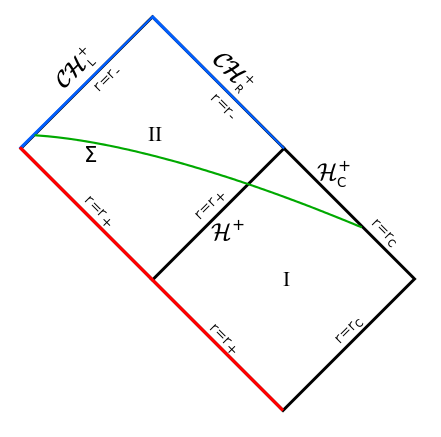}
	\caption{Penrose diagram for Kerr-de Sitter. The event and cosmological horizons are $\mathcal{H}^+$ and $\mathcal{H}_C^+$, respectively, the blue lines are the left and right Cauchy horizons $\mathcal{CH}_{L,R}^+$, the green line is a spacelike hypersurface extending from the cosmological horizon across the event horizon and the left Cauchy horizon. Quasinormal modes blow up along the white hole horizon (red line) and also along the past cosmological horizon.}
	\label{fig:penrose}
\end{figure}

The behaviour of perturbations at the Cauchy horizon depends on the rate of decay of perturbations along the event horizon \cite{Hintz:2015jkj}. Faster decay along the event horizon gives a milder instability of the Cauchy horizon. With positive cosmological constant, it has been proved that perturbations decay {\it exponentially} along the event horizon. Specifically, for massless scalar field perturbations of Reissner-Nordstr\"om-de Sitter, or slowly rotating Kerr-de Sitter, black holes it has been proved  that, there exist constants $\Phi_0$ and $C,\alpha>0$ such that, outside the black hole \cite{Barreto:1997,BoHa:2008,Dyatlov:2011jd,Dyatlov:2013hba,Hintz:2016gwb,Hintz:2016jak}
\be
\label{gap}
 |\Phi - \Phi_0| \le C e^{-\alpha t}
\ee
where $t$ labels a foliation by spacelike hypersurfaces that extend from the future event horizon to the future cosmological horizon (e.g. the surface $\Sigma$ of Fig. \ref{fig:penrose}), with the hypersurfaces related by the time translation symmetry of the black hole. The constant $\alpha$ is called the spectral gap. The spectral gap can be determined by looking at the most slowly decaying {\it quasinormal modes} of the black hole: $\alpha$ is the largest number such that $\alpha\le-{\rm Im}(\omega)$ for all quasinormal frequencies $\omega$. 

If $\alpha$ is known then one can determine the behaviour of generic perturbations at the Cauchy horizon and hence ascertain whether or not strong cosmic censorship is violated. And $\alpha$ can be determined by looking at quasinormal modes of the black hole. This is what was done in Ref.  \cite{Cardoso:2017soq} for Reissner-Nordstr\"om-de Sitter black holes. By determining (numerically) the most slowly decaying quasinormal modes, the value of $\alpha$ was determined. For black holes sufficiently close to extremality, the value of $\alpha$ was sufficiently large to indicate that, when nonlinearities are included (e.g. using results of Ref. \cite{costa}), it would be possible to extend the solution across the Cauchy horizon as a weak solution of the equations of motion, in violation of the strong cosmic censorship conjecture.

Reissner-Nordstr\"om-de Sitter black holes are not very relevant physically. However, they are often viewed as a toy model for the much more physical case of Kerr-de Sitter black holes. The massless scalar field can be viewed as a toy model for linearized gravitational perturbations. So the results of Ref. \cite{Cardoso:2017soq} suggest that maybe there is a violation of strong cosmic censorship for nearly extremal Kerr-de Sitter black holes in vacuum. Indeed this was conjectured in Ref. \cite{Dafermos:2012np}. That is what we will investigate in this paper. 
 
Our approach is the following. We will study linear perturbations of a non-extremal Kerr-de Sitter black hole. These perturbations could be either a massless scalar field or linearized gravitational perturbations. Such a linear perturbation will source a second order metric perturbation. The linear perturbation will be continuous but not necessarily differentiable at the Cauchy horizon. However, in order to extend beyond the Cauchy horizon, the linear solution needs to be sufficiently regular that the equation of motion for the second order perturbation can be satisfied in a weak sense at the Cauchy horizon. As we will explain, this leads to the criterion that the scalar field, or linearized metric perturbation, must have a locally square integrable derivative, \emph{i.e.}, it should belong to the Sobolev space $H^1_{\rm loc}$. This was also the criterion used in Ref. \cite{Cardoso:2017soq}.

Consider a scalar field quasinormal mode in a non-extremal Kerr-de Sitter spacetime. Such a solution has definite frequency and satisfies ingoing boundary conditions at the future event horizon  $\mathcal{H}^+$, and outgoing boundary conditions at the future cosmological horizon  $\mathcal{H}_C^+$ (see Fig. \ref{fig:penrose}). Working in coordinates regular across ${\cal H}^+$, a quasinormal mode can be analytically continued into the black hole interior (region II of Fig. \ref{fig:penrose}). We will determine how such a quasinormal mode behaves at the Cauchy horizon $\mathcal{CH}_{R}^+$ of Fig. \ref{fig:penrose}. It is straightforward to show that it belongs to $H^1_{\rm loc}$ if, and only if, minus the imaginary part of the quasinormal frequency exceeds a certain value, \emph{i.e.}, the mode decays fast enough. 

We will use geometric optics and numerics to show that there always exist ``photon sphere" quasinormal modes whose decay is slow enough that, when continued inside the black hole, they do not belong to $H^1_{\rm loc}$ at the Cauchy horizon $\mathcal{CH}_{R}^+$. We can now prove that strong cosmic censorship is respected as follows. Assume that one is given initial data on the surface $\Sigma$ shown in Fig. \ref{fig:penrose}, for a linearized perturbation which belongs to $H^1_{\rm loc}$ at $\mathcal{CH}_{R}^+$. Now ``perturb this perturbation" by adding the initial data for our quasinormal mode, with an arbitrary amplitude. This produces a new perturbation which does not belong to $H^1_{\rm loc}$ at $\mathcal{CH}_{R}^+$. Hence a {\it generic} perturbation does not belong to $H^1_{\rm loc}$ and so it cannot be extended beyond $\mathcal{CH}_{R}^+$ consistently with the equations of motion. Hence strong cosmic censorship is respected.\footnote{Note that we do not need to assume the validity of equation (\ref{gap}), which is just as well because (\ref{gap}) has been established only for slowly rotating black holes.} 

For linearized gravitational perturbations, we exploit the fact that there exist a gauge invariant component of the Weyl tensor which satisfies a decoupled equation of motion. If the linearized metric perturbation belongs to $H^1_{\rm loc}$ in some gauge then the blow up of this Weyl component at ${\cal CH}_R^+$ cannot exceed a certain rate\footnote{The precise condition is that, in a regular tetrad, this Weyl component must belong to the Sobolev space $H^{-1}_{\rm loc}$.}. However, for some photon sphere quasinormal modes we find that the blow up does exceed this rate. This proves that there exists no gauge in which the linearized metric perturbation is in $H^1_{\rm loc}$. This proves that strong cosmic censorship is respected by gravitational perturbations of any non-extremal Kerr-de Sitter black hole. 

{\bf Note added.} As this paper was nearing completion, we received Ref. \cite{Hod:2018dpx}. This paper considers perturbations of Reissner-Nordstr\"om-de Sitter black holes by a scalar field which is {\it charged} and has non-zero mass. It was argued that, for sufficiently large charge and mass, the decay of such a field is always sufficiently slow to ensure that strong cosmic censorship is respected.

%%%%%%%%%%%%%%%%%%%%%%%%%%%%%%%%%%%%%%%%%%%%%%%%%%%%%%%%%%%%
%%%%%%%%%%%%%%%%%%%%%%%%%%%%%%%%%%%%%%%%%%%%%%%%%%%%%%%%%%%%
\section{Weak solutions}
%%%%%%%%%%%%%%%%%%%%%%%%%%%%%%%%%%%%%%%%%%%%%%%%%%%%%%%%%%%%
%%%%%%%%%%%%%%%%%%%%%%%%%%%%%%%%%%%%%%%%%%%%%%%%%%%%%%%%%%%%
We will be discussing linear perturbations which are continuous, but not necessarily differentiable, at the Cauchy horizon. The fundamental question that needs to be addressed is whether there is any sense in which such a perturbation can satisfy the equations of motion at the Cauchy horizon. Moreover, we are primarily interested in answering this question for {\it nonlinear} perturbations. We will explain why this leads to the condition that linear perturbations should belong to $H^1_{\rm loc}$. 

Consider a scalar field $\Phi$ satisfying $\Box \Phi=0$. Treat this as a first order perturbation, sourcing a second order metric perturbation $h^{(2)}_{\mu\nu}$. Then $h^{(2)}_{\mu\nu}$ will satisfy
\be
\label{h2eq}
 {\cal L} h^{(2)}_{\mu\nu} = 8 \pi T_{\mu\nu}[\Phi]\,,
\ee 
where ${\cal L}$ is a certain second order differential operator and $T_{\mu\nu}[\Phi]$ is the energy momentum of the scalar field. Now assume that $\Phi$ and $h^{(2)}_{\mu\nu}$ are not necessarily continuously differentiable. One can still make sense of the above equation by multiplying by a smooth, compactly supported, symmetric tensor, $\psi^{\mu\nu}$ and integrating by parts:\footnote{The test function $\psi^{\mu\nu}$ permits integrating by parts without introducing boundary terms.}
\be
 \int d^4 x \sqrt{-g} \left( h^{(2)}_{\mu\nu} {\cal L}^\dagger \psi^{\mu\nu}   - 8 \pi \psi^{\mu\nu} T_{\mu\nu}  \right)  =0\,,
\ee
where ${\cal L}^\dagger$ is the adjoint of ${\cal L}$ arising from the integration by parts. 
If this equation is satisfied for {\it any} smooth, compactly supported symmetric $\psi^{\mu\nu}$ then we say that we have a {\it weak solution} of (\ref{h2eq}). In order for this equation to make sense, the terms involving the scalar field must be finite, which is guaranteed by demanding that $\Phi$ belongs to $H^1_{\rm loc}$. This is the space of functions $\Phi$ defined by the condition that, for any smooth compactly supported function $\psi$, the quantity $( \hat{\Phi}^2 +\partial_\mu \hat{\Phi} \partial_\mu \hat{\Phi})$ is integrable, where $\hat{\Phi} \equiv \psi \Phi$.

Similarly, if one starts from a linearized gravitational perturbation $h_{\mu\nu}$  one can consider the second order perturbation $h^{(2)}_{\mu\nu}$ sourced by the linear perturbation. This satisfies an equation analogous to (\ref{h2eq}) where the RHS is quadratic in first derivatives of $h_{\mu\nu}$. So repeating the above argument, the minimum regularity required of $h_{\mu\nu}$ in order for the equation for $h^{(2)}_{\mu\nu}$ to be satisfied weakly is that, in some gauge, $h_{\mu\nu}$ should belong to $H^1_{\rm loc}$. 

We can relate this to the criterion for weak solutions of the full nonlinear vacuum Einstein equation. Applying the above procedure to the Einstein equation results in the criterion that, in some chart, the Christoffel symbols should be locally square integrable \cite{Christodoulou:2008nj}. In such a chart, perform a perturbative expansion of the metric $g_{\mu\nu} = \bar{g}_{\mu\nu} + h_{\mu\nu} + h^{(2)}_{\mu\nu} + \ldots$ and consider the integral of the sum of squares of the Christoffel symbols. At first order this will give terms linear in $h_{\mu\nu}$ and its first derivative. So at first order the minimum regularity required is that $h_{\mu\nu}$ and its first derivative be integrable. However, at second order, terms quadratic in first derivatives $h_{\mu\nu}$ will arise, and so we will need first derivatives of $h_{\mu\nu}$ to be square integrable and hence we will need $h_{\mu\nu}$ to belong to $H^1_{\rm loc}$. Continuing to higher orders does not give anything new because all terms are at most quadratic in first derivatives of $h_{\mu\nu}$. %%%%%%%%%%%%%%%%%%%%%%%%%%%%%%%%%%%%%%%%%%%%%%%%%%%%%%%%%%%%%%%%%%%%%%%%%
%%%%%%%%%%%%%%%%%%%%%%%%%%%%%%%%%%%%%%%%%%%%%%%%%%%%%%%%%%%%
\section{Kerr-de Sitter}
%%%%%%%%%%%%%%%%%%%%%%%%%%%%%%%%%%%%%%%%%%%%%%%%%%%%%%%%%%%%
%%%%%%%%%%%%%%%%%%%%%%%%%%%%%%%%%%%%%%%%%%%%%%%%%%%%%%%%%%%%
\subsection{Coordinates}
%%%%%%%%%%%%%%%%%%%%%%%%%%%%%%%%%%%%%%%%%%%%%%%%%%%%%%%%%%%%
We will write the Kerr-de Sitter metric \cite{Carter:1968ks} as follows \cite{Chambers:1994ap} 
\be
\label{metric}
 ds^2 = \rho^2 \left[\frac{\mathrm{d}r^2}{\Delta_r} + \frac{\mathrm{d}\chi^2}{\Delta_\chi} \right] + \frac{1}{\rho^2\Xi^2}\left[\Delta_\chi \left(\mathrm{d}t - \frac{ \sigma_r}{a} \mathrm{d}\phi\right)^2 - \Delta_r \left(\mathrm{d}t - \frac{\sigma_\chi}{a} \mathrm{d}\phi\right)^2\right]
\ee
where
\be
 \sigma_r = a^2+r^2\,, \qquad \sigma_\chi =  a^2-\chi^2\,, \qquad \rho^2 = r^2 + \chi^2\,, \qquad \Xi = 1 + \frac{a^2}{L^2}
\ee
and
\be
\Delta_r =  \sigma_r\left(1 - \frac{r^2}{L^2} \right) - 2\,M\,r\,, \qquad \Delta_\chi = \sigma_\chi \left(1 + \frac{\chi^2}{L^2} \right)\,, \qquad \Lambda=\frac{3}{L^2}\,.
\ee
In these coordinates, $\phi\in[0,2\pi)$ and $\chi\in[-|a|,|a|]$. It is convenient to define
\be
 \Omega(r) = \frac{a}{r^2+a^2}\,.
\ee
We assume that the solution describes a non-extremal black hole, which implies that there are three real positive roots of $\Delta_r$, satisfying $r_-<r_+<r_c$. These correspond to the Cauchy horizon, event horizon and cosmological horizon, respectively. The angular velocities of the horizons will be denoted by
\be
 \Omega_- = \Omega(r_-), \qquad \Omega_+ = \Omega(r_+), \qquad \Omega_c = \Omega(r_c).
\ee
Starting from the above metric with $r_+<r<r_c$, which we call region I (see Fig. \ref{fig:penrose}), we define ingoing coordinates $(v,r,\chi,\phi')$ as follows:
\be
\label{vphi'}
 \mathrm{d}t =  \mathrm{d}v - \frac{\Xi\,\sigma_r}{\Delta_r}  \mathrm{d}r \qquad  \mathrm{d}\phi =  \mathrm{d}\phi' - \frac{a\,\Xi}{\Delta_r}  \mathrm{d}r \,.
\ee
In the ingoing coordinates, we can extend across $r=r_+$ into a new region, region II (see Fig. \ref{fig:penrose}), with $r_-<r<r_+$. In the new coordinates $g_{rr}=0$ so $\partial/\partial r$ is globally null. In fact $\partial/\partial r$ is also geodesic and shear free: it is one of the repeated principal null directions of the solution; $-\partial/\partial r$ is tangent to ingoing null geodesics. 

In region II we can re-introduce the original coordinates $(t,r,\chi,\phi)$ using (\ref{vphi'}). The metric in these coordinates takes the same form as \eqref{metric}. Now, in region II, we introduce outgoing coordinates $(u,r,\chi,\phi'')$ defined by
\be
 \label{uphi''}
  \mathrm{d}t =  \mathrm{d}u + \frac{\Xi\,\sigma_r}{\Delta_r}  \mathrm{d}r \qquad  \mathrm{d}\phi =  \mathrm{d}\phi''+  \frac{a \,\Xi}{\Delta_r} \mathrm{d} r \,.
\ee
This lets us analytically continue the metric across the ``right" Cauchy horizon $\mathcal{CH}_{R}^+$ in region II into a new region with $r<r_-$. In these coordinates, $-\partial/\partial r$ is null, geodesic and shear free, and future-directed. It is the second repeated principal null direction of the solution. It is tangent to outgoing null geodesics in region II, \emph{i.e.}, null geodesics which cross ${\cal CH}_R^+$. 

\begin{figure}
\centering
    \includegraphics[width=0.45\textwidth]{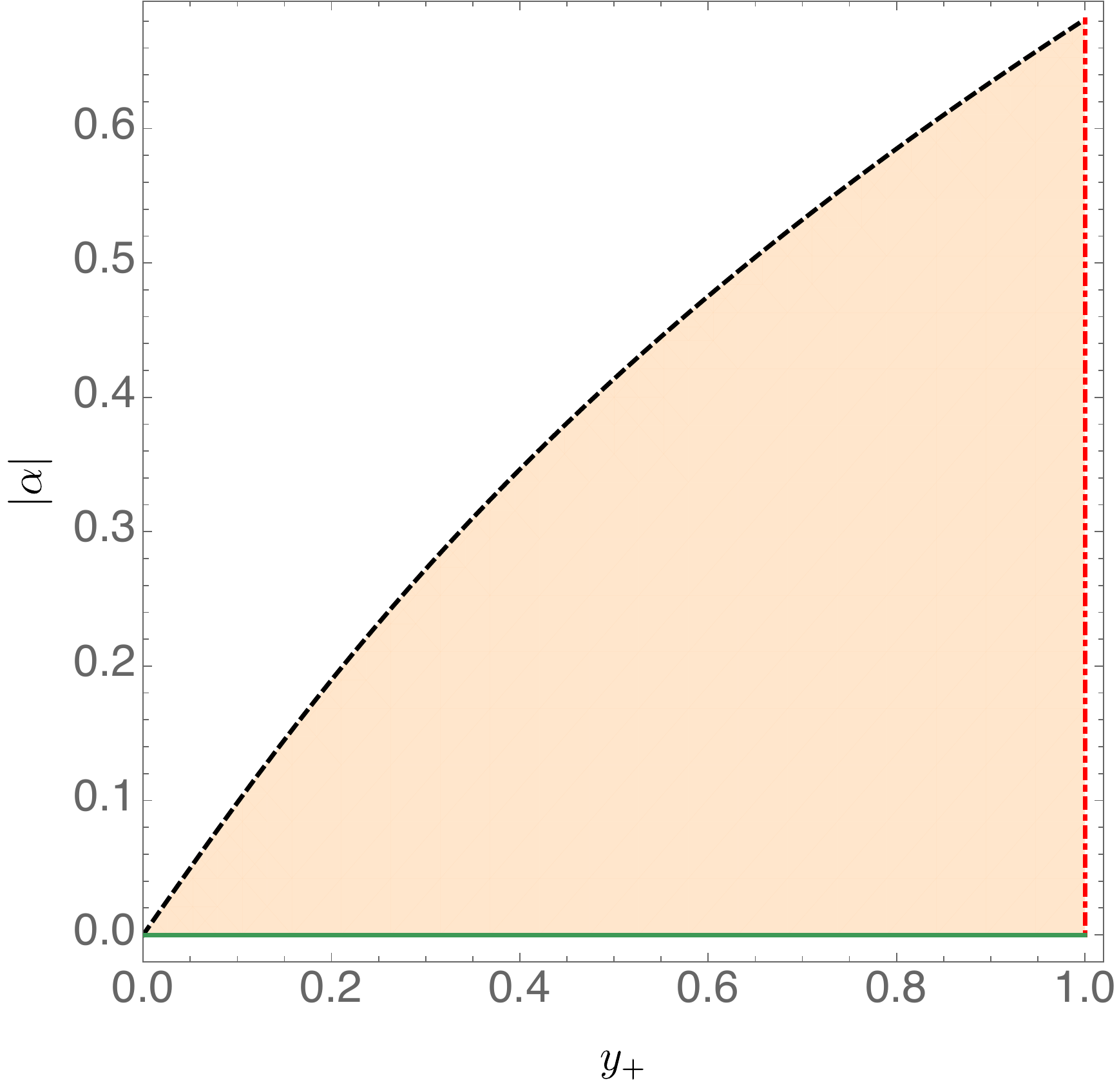}
  \caption{Moduli space of solutions in the $(y_+,\alpha)$ plane: the dashed black curve corresponds to extremality where $r_+=r_-$, the red dotted-dashed line corresponds to the limit where the black hole horizon coincides with the cosmological horizon and the green solid line to a Schwarzschild-de Sitter black hole.}
\label{fig:moduli}
\end{figure}

We will parametrise Kerr-de Sitter solutions using the dimensionless quantities 
\be
\{y_+,\alpha\}\equiv\{r_+/r_c,a/r_c\}
\ee
These variables are in one-to-one correspondence with members of the Kerr-de Sitter family of solutions and mean we essentially normalise all our quantities to $r_c$. The moduli space of solutions is shown in Fig.~\ref{fig:moduli}. Kerr-de Sitter black holes have three distinct extremal limits: $r_+=r_-$, $r_+=r_c$ and $r_+=r_-=r_c$. The first two are marked as the black dashed line and red dotted-dashed line in Fig.~\ref{fig:moduli}, respectively. For completeness, we also show in Fig.~\ref{fig:moduli} the Schwarzschild limit marked as a green solid line.
When $r_+=r_-$, we have
\be
|a|=|a_{\mathrm{ext}}|\equiv\frac{r_c}{\sqrt{2}}\sqrt{(1+y_+)\sqrt{1+2y_++9y_+^2}-y_+(2+3y_+)-1}\,.
\ee
%%%%%%%%%%%%%%%%%%%%%%%%%%%%%%%%%%%%%%%%%%%%%%%%%%%%%%%%%%%%
\subsection{Tetrad}
%%%%%%%%%%%%%%%%%%%%%%%%%%%%%%%%%%%%%%%%%%%%%%%%%%%%%%%%%%%%
When we study gravitational perturbations of Kerr-de Sitter black holes, it will be useful to introduce a null tetrad $\{\NPl,\NPn,\NPm,\bar{\NPm}\}$ satisfying the following orthogonality relations
\be
\NPl \cdot \NPn=-1\,,\quad \bar{\NPm}\cdot \NPm=1
\ee
with all remaining combinations of inner products giving zero, and $g_{\mu\nu}=-2\NPl_{(\mu}\NPn_{\nu )}+2\NPm_{(\mu}\NPbm_{\nu )}$.

There is obviously a lot of freedom in choosing such tetrad, and some choices make the equations governing gravitational perturbations of Kerr-de Sitter black holes easier than others. Here we will choose the Chambers-Moss null tetrad $\{\NPl,\NPn,\NPm,\NPbm\}$ \cite{Chambers:1994ap}, which in $\{t,r,\chi,\phi\}$ coordinates, reads:
\begin{eqnarray}
 &&\NPl^{\mu}\partial_{\mu}=\frac{1}{\sqrt{2}\sqrt{r^2+\chi ^2}}\left(\Xi  \,\frac{a^2+r^2}{\sqrt{\Delta _r}} \,\partial _t+\sqrt{\Delta _r} \,\partial _r+\frac{a  \,\Xi  }{\sqrt{\Delta _r}} \,\partial _{\phi }\right),  \nonumber\\
 && \NPn^{\mu}\partial_{\mu}=\frac{1}{\sqrt{2}\sqrt{r^2+\chi ^2}}\left(\Xi   \,\frac{a^2+r^2}{\sqrt{\Delta _r}} \,\partial _t-\sqrt{\Delta _r} \,\partial _r+ \frac{a  \,\Xi }{\sqrt{\Delta _r}} \,\partial _{\phi }\right),   \nonumber\\
 &&  \NPm^{\mu}\partial_{\mu}= -\frac{i}{\sqrt{2}\sqrt{r^2+\chi ^2}}\left( \Xi  \,\frac{a^2-\chi ^2}{\sqrt{\Delta _{\chi }}} \,\partial _t+ i\sqrt{\Delta _{\chi }} \,\partial _{\chi }+\frac{a  \,\Xi }{\sqrt{\Delta _{\chi }}} \,\partial _{\phi }\right)\,,
 \label{KerrdS:NPtetrad}
\end{eqnarray}
and $\NPbm$ is the complex conjugate of $\NPm$.

We will need to investigate the regularity of such a tetrad across the Cauchy horizon. So we need to write it in outgoing coordinates $\{u,r,\chi,\phi^{\prime\prime}\}$:
\begin{eqnarray}
 &&\NPl^{\mu}\partial_{\mu}=\frac{\sqrt{\Delta_r}}{\sqrt{2}\sqrt{r^2+\chi^2}}\partial_r,  \nonumber\\
 && \NPn^{\mu}\partial_{\mu}=\frac{\sqrt{2}}{\sqrt{r^2+\chi ^2}}\left(\Xi   \,\frac{a^2+r^2}{\sqrt{\Delta _r}} \,\partial _u-\frac{\sqrt{\Delta _r}}{2} \,\partial _r+ \frac{a  \,\Xi }{\sqrt{\Delta _r}} \,\partial _{\phi^{\prime\prime} }\right),   \nonumber\\
 &&  \NPm^{\mu}\partial_{\mu}= -\frac{i}{\sqrt{2}\sqrt{r^2+\chi ^2}}\left( \Xi  \,\frac{a^2-\chi ^2}{\sqrt{\Delta _{\chi }}} \,\partial _u+ i\sqrt{\Delta _{\chi }} \,\partial _{\chi }+\frac{a  \,\Xi }{\sqrt{\Delta _{\chi }}} \,\partial _{\phi^{\prime\prime} }\right)\,.
 \label{KerrdS:NPtetradout}
\end{eqnarray}
However the tetrad (\ref{KerrdS:NPtetradout}) is not regular when $\Delta_r=0$ (for instance at the Cauchy horizon $r=r_-$) since $\NPn$ blows up there. To fix this, we change to a new tetrad where
\begin{align}
 \widetilde{\NPl}=\frac{1}{\sqrt{\Delta_r}}\NPl\,, \qquad \widetilde{\NPn}=\sqrt{\Delta_r}\NPn  \qquad\text{and} \qquad \widetilde{\NPm}=\NPm\,,  \label{KerrdS:NPtetradoutSmooth}
\end{align}
which is now smooth when $\Delta_r=0$.

%%%%%%%%%%%%%%%%%%%%%%%%%%%%%%%%%%%%%%%%%%%%%%%%%%%%%%%%%%%%
%%%%%%%%%%%%%%%%%%%%%%%%%%%%%%%%%%%%%%%%%%%%%%%%%%%%%%%%%%%%
\section{Scalar field quasinormal modes}
%%%%%%%%%%%%%%%%%%%%%%%%%%%%%%%%%%%%%%%%%%%%%%%%%%%%%%%%%%%%
%%%%%%%%%%%%%%%%%%%%%%%%%%%%%%%%%%%%%%%%%%%%%%%%%%%%%%%%%%%%
\subsection{Preliminaries}
%%%%%%%%%%%%%%%%%%%%%%%%%%%%%%%%%%%%%%%%%%%%%%%%%%%%%%%%%%%%
Consider a scalar field $\Phi$ obeying the wave equation $\Box \Phi=0$. Quasinormal modes are solutions of the following form 
\be
\label{Phi_ans}
 \Phi = e^{-i\omega t} e^{i m \phi} S_{\omega \ell m}(\chi) R_{\omega \ell m}(r)
\ee
where $\ell = 0,1,2, \ldots$, $|m| \le \ell$ and the frequency $\omega$ is determined in terms of $\ell,m$ and an ``overtone" number $n=0,1,2,\ldots$. These quasinormal frequencies are determined by the condition that the solution obeys ingoing boundary conditions as $r\rightarrow r_+$ and outgoing boundary conditions as $r \rightarrow r_c$, \emph{i.e.}, the solution is smooth at the future event horizon $\mathcal{H}^+$ and at the future cosmological horizon $\mathcal{H}_C^+$. Quasinormal frequencies are complex: $\omega = \omega_R + i \omega _I$ with $\omega_I <0$ so quasinormal modes decay exponentially with time outside the black hole. 

If we use ingoing coordinates $(v,r,\chi,\phi')$, regular in regions I and II of Fig. \ref{fig:penrose}, then a quasinormal mode is an analytic function of the coordinates in region I and can be analytically continued into region II. In the ingoing coordinates, a quasinormal mode has time dependence $e^{-i \omega v}$, so it will diverge as $v \rightarrow - \infty$, \emph{i.e.}, along the red line on Fig. \ref{fig:penrose}. We are interested in the behaviour of the mode at the Cauchy horizon $\mathcal{CH}_{R}^+$. To investigate regularity there we need to convert to outgoing coordinates in the black hole interior. 

In region II, we can convert from the ingoing coordinates to coordinates $(t,r,\chi,\phi)$ and the quasinormal mode will again take the form (\ref{Phi_ans}). Now converting (\ref{Phi_ans}) to outgoing coordinates $(u,r,\chi,\phi'')$ in region II gives 
\be
 \Phi = e^{-i\omega u} e^{i m \phi''} S_{\omega \ell m}(\chi) \tilde{R}_{\omega \ell m}(r)
\ee
for some function $\tilde{R}_{\omega \ell m}$. Near the right Cauchy horizon $\mathcal{CH}_{R}^+$, there are two independent solutions of this form, which behave as follows
\begin{subequations}
\begin{align}
& \Phi^{(1)} = e^{-i \omega u} e^{i m \phi''} S_{\omega \ell m}(\chi) \hat{R}^{(1)}_{\omega \ell m}(r)\,, \\
& \Phi^{(2)} = e^{-i \omega u} e^{i m \phi''} S_{\omega \ell m}(\chi) (r-r_-)^{i (\omega - m \Omega_-)/\kappa_-}  \hat{R}^{(2)}_{\omega \ell m}(r)\,,
\end{align}
\end{subequations}
where $\hat{R}^{(1,2)}$ denote smooth functions which are non-zero at $r=r_-$, and $\Omega_- = \Omega(r_-)$. Notice that ${\rm Im}(\omega)<0$ implies that $\Phi^{(2)}$ vanishes at $r=r_-$. However $\Phi^{(2)}$ is not smooth at $r=r_-$.  At the Cauchy horizon, our quasinormal mode will be some linear combination of the above two solutions. There is no reason why either of the coefficients in this linear combination should vanish. Hence the regularity of the quasinormal mode is determined by the non-smooth solution $\Phi^{(2)}$. What is the condition for $\Phi^{(2)}$ to be locally square integrable? We have $\Phi^{(2)} \sim (r-r_-)^p$ with $p = i(\omega - m \Omega_-)/\kappa_-$. Hence $\partial_r \Phi^{(2)} \sim (r-r_-)^{p-1}$ which is square integrable if, and only if, $2(\beta -1)>-1$ where $\beta = {\rm Re}(p)$. In other words the condition for our quasinormal mode to belong to $H^1_{\rm loc}$ at the Cauchy horizon is
\be
 \beta > \frac{1}{2} \qquad {\rm where} \qquad \beta \equiv -\frac{{\rm Im}(\omega)}{\kappa_-}\,.
\ee 
Therefore if we can find a quasinormal mode with $\beta <1/2$ then the scalar field cannot be extended across the Cauchy horizon in $H^1_{\rm loc}$ and so strong cosmic censorship is respected. On the other hand if {\it all} quasinormal modes have $\beta>1/2$ then strong cosmic censorship may be violated. Ref. \cite{Cardoso:2017soq} argued that the latter is what happens for nearly extremal Reissner-Nordstr\"om-de Sitter black holes.

%%%%%%%%%%%%%%%%%%%%%%%%%%%%%%%%%%%%%%%%%%%%%%%%%%%%%%%%%%%%
\subsection{\label{sec:geo}Geometric optics}
%%%%%%%%%%%%%%%%%%%%%%%%%%%%%%%%%%%%%%%%%%%%%%%%%%%%%%%%%%%%

In the eikonal limit, also known as geometric optics limit, where $\ell\gg 1$ ($\ell\sim |m| \gg 1$ for spinning backgrounds) there are quasinormal mode frequencies $-$ known as ``photon sphere" quasinormal modes $-$  which are related to the properties of the unstable circular photon orbits in the equatorial plane. Namely, the real part $\omega_R$ of the frequency is proportional to the Keplerian frequency $\Omega_c$ of the circular null orbit and the imaginary part of the frequency is proportional to the Lyapunov exponent $\lambda$ of the orbit \cite{Goebel:1972,Ferrari:1984zz,Ferrari:1984ozr,Mashhoon:1985cya,Bombelli:1991eg,Cornish:2003ig,Cardoso:2008bp,Dolan:2010wr,Yang:2012he,stuchlik}. The latter describes how quickly a null geodesic congruence on the circular orbit increases its cross section under infinitesimal radial deformations. 

These photon sphere quasinormal modes turn out to play a fundamental role in our discussion. Therefore, in this section we will use geometric optics to compute these modes for the Kerr-de Sitter background. In the next section we will find that the resulting analytical expression for the frequency matches extremely well the values that we find numerically already for values of $\ell=m$ as low as 10.

The geodesic equation, describing the motion of point-like particles around a Kerr-de Sitter black hole, is known to lead to a set of quadratures. This is perhaps an unexpected result, since Kerr-de Sitter only possesses two Killing fields, given in our coordinate system as $K=\partial/\partial_t$ and $M=\partial/\partial_\phi$ and thus seems one short of leading to an integrable system. However, there is another conserved quantity, the Carter constant, associated to a Killing tensor $K_{ab}$, which saves the day \cite{ChandrasekharBook}.

The most direct way to see this integrable structure is to look at the Hamilton-Jacobi equation \cite{ChandrasekharBook}:
\be
\frac{\partial S}{\partial x^\mu}\frac{\partial S}{\partial x^\nu}g^{\mu\nu}=0\,,
\ee
where $S$ is known as the principal function.  One can recover the motion of null particles by noting that, according to Hamilton-Jacobi's theory,
\be
\frac{\partial S}{\partial x^\mu}\equiv p_\mu\quad \text{and}\quad p^\mu =\frac{\mathrm{d}x^\mu}{\mathrm{d}\tau}\,,
\label{eq:def}
\ee
with $\tau$ denoting an affine parameter.

We then take a separation \emph{ansatz} of the form
\be
S=-e\,t+j\,\phi+R(r)+X(\chi)\,,
\label{eq:sep}
\ee
which gives the following coupled ordinary differential equations for $R(r)$ and $X(\chi)$
\begin{subequations}
\begin{align}
& \Delta_r^2(\partial_r R)^2-\Xi^2\left(e\sigma_r-a j\right)^2+\left[\mathcal{Q}+\Xi^2(j-ae)^2\right]\Delta_r=0\,,
\\
& \Delta_\chi^2(\partial_\chi X)^2-\Xi^2\left(e\sigma_\chi-a j\right)^2-\left[\mathcal{Q}+\Xi^2(j-ae)^2\right]\Delta_\chi=0\,,
\end{align}
\label{eqs:geo}
\end{subequations}
where $\mathcal{Q}$ is a separation constant known as the Carter constant. The constants $e$ and $j$ are the conserved charges associated with the Killing fields $K$ and $M$\footnote{For massive particles, these coincide with the energy and angular momentum of the particle, but for massless particles $e$ and $j$ have no physical meaning since they can be rescaled. The ratio $j/e$, however, is invariant under such rescallings.} via
\be
e\equiv - K_\mu \dot{x}^\mu\qquad \text{and}\qquad j\equiv M_\mu \dot{x}^\mu\,.
\label{eq:conserved}
\ee

Eqs.~(\ref{eqs:geo}) translate into a statement about the particle trajectories via ~\eqref{eq:def} and \eqref{eq:sep}. In particular, for $\dot{\chi}$, we find
\begin{equation}
(r^2+\chi^2)^2\frac{\dot{\chi}^2}{\Delta_\chi}=\mathcal{Q}-e^2\Xi^2\left[\frac{(a b-\sigma_\chi)^2}{\Delta_\chi}-(b-a)^2\right],
\end{equation}
where we define the geodesic impact parameter by
\be
b\equiv\frac{j}{e}\,.
\ee

Since we are interested in matching the behaviour of geodesics with that of quasinormal modes with large values of $\ell=m$, we can restrict attention to the equatorial plane for which $\chi=0$. This can only be the case if initially $\chi(0)=\dot{\chi}(0)=0$ and $\mathcal{Q}=0$. The equation governing the radial motion now gives
\be
\label{eq:geodesic}
\dot{r}^2=V(r;b)\,,
\ee
where
\be
V(r;b)=\frac{j^2\Xi^2}{b^2}\left\{1+\frac{(a-b)^2}{L^2}+\frac{(a-b)}{r^2}\left[a+b+\frac{a^2}{L^2}(a-b)\right]+\frac{2 M(a-b)^2}{r^3}\right\}\,.
\label{eq:pot}
\ee

We are now interested in finding the photon sphere (region where null particles are trapped on circular unstable orbits), \emph{i.e.} the values of $r=r_s$ and $b=b_s$, such that
\be
V(r_s,b_s)=0\quad\text{and}\quad \left.\partial_r V(r,b)\right|_{r=r_s,b=b_s}=0.
\ee
From the second equation above we get
\be
b_s(r_s)=a\,\frac{a^2 r_s+L^2 \left(3 M+r_s\right)}{a^2 r_s+L^2 \left(3 M-r_s\right)}\,,
\ee
while from the first we get:
\be
a^4 r_s^3+a^2 \left[2 L^2 r_s^2 \left(3 M+r_s\right)-4 L^4 M\right]+L^4 r_s \left(r_s-3 M\right)^2=0\,.
\ee
The two relevant real roots, \emph{i.e.} those that satisfy $r_+\leq r_s^{\pm}\leq r_c$, can be written as
\begin{equation}
r_s^{\pm} = \frac{2 M}{\Xi ^2} \left\{\gamma^-+\gamma \cos \left[\frac{2}{3} \arccos\left(\mp  \sqrt{\frac{1}{2}-\frac{\gamma^-\gamma^+}{2\gamma^{3}}+\frac{a^2 \Xi ^4}{M^2\gamma^{3}}}\right)\right]\right\}
\label{eq:orbits}
\end{equation}
where
\be
\gamma\equiv \sqrt{1-\frac{14 a^2}{L^2}+\frac{a^4}{L^4}}\,,\quad \gamma^+\equiv1+\frac{34 a^2}{L^2}+\frac{a^4}{L^4}\quad \text{and}\quad \gamma^-\equiv1-\frac{a^2}{L^2}\,.
\ee
At first glance, it might appear that the argument of the square root appearing as the argument of the $\arccos $ in Eq.~(\ref{eq:orbits}), as well as the definitions above, might become negative. However, we have explicitly checked that this is not the case whenever the line element (\ref{metric}) describes a black hole. This is consistent with Ref. \cite{stuchlik}, which argues that a Kerr de Sitter black hole always has two circular photon orbits. The signs are chosen such that $r_s^+$ corresponds to prograde orbits, \emph{i.e.} $b^+_s\equiv b_s(r_s^+)>0$ and $r_s^-$ to retrograde orbits, \emph{i.e.} $b^-_s\equiv b_s(r_s^-)<0$.

We can now compute the orbital angular velocity (aka Kepler frequency) of our null circular photon orbit, which is simply given by
\be
\Omega_c^{\pm}\equiv \frac{\dot{\phi}}{\dot{t}}=\frac{1}{b_s^{\pm}}\,,
\ee
where in the second equality we have used Eq.~(\ref{eq:conserved}) and took $r=r_s^\pm$ and $b=b_s^\pm$.

On an orbit with impact parameter $b=b_s^{\pm}$, the radial potential (\ref{eq:pot}) simplifies considerably,
\be
V(r;b_{s}^{\pm})=\frac{j^2\,\Xi^2}{(b_s^{\pm})^2}\,(\beta_s^\pm)^2\left(1-\frac{r_s^\pm}{r}\right)^2\left(1+\frac{2\,r^{\pm}_s}{r}\right)\,,
\label{eq:potsimple}
\ee
where we defined
\be
(\beta_s^\pm)^2=1+\frac{\left(a-b_s^{\pm}\right)^2}{L^2}\,.
\ee

The final step in our calculation is to compute the largest Lyapunov exponent $\lambda$, measured in units of $t$, associated with infinitesimal fluctuations around photon orbits with $r(\tau)=r_s^\pm$. This can be readily done by perturbing the geodesic equation (\ref{eq:geodesic}) with the simplified potential (\ref{eq:potsimple}) and setting $r(\tau)=r_s^\pm+\delta r(\tau)$. One finds that small deviations obey
\be
\delta r(t)=\mathrm{exp}\left[+\frac{\sqrt{3}}{\Xi}\,\beta_s^\pm\,\frac{a^2-a \,b_s^\pm+(r_s^\pm)^2}{(b_s^{\pm}-a)\,b_s^\pm \,r_s^\pm}\,t\right]+C_+\,,
\ee
and
\be
\delta r(t)=\mathrm{exp}\left[- \frac{\sqrt{3}}{\Xi}\,\beta_s^\pm\,\frac{a^2-a \,b_s^\pm+(r_s^\pm)^2}{(b_s^{\pm}-a)\,b_s^\pm \, r_s^\pm}\,t\right]+C_-\,,
\ee
where $C_{\pm}$ are integration constants. The largest Lyapunov exponent is simply given by
\be
\lambda^{\pm}=\left|\frac{\sqrt{3}}{\Xi}\,\beta_s^\pm\,\frac{a^2-a \,b_s^\pm+(r_s^\pm)^2}{(b_s^{\pm}-a)\,b_s^\pm \,r_s^\pm}\right|\,.
\ee

One reconstructs the approximate spectrum of the photon sphere family of quasinormal modes with $\ell=|m|\gg 1$ using \cite{Goebel:1972,Ferrari:1984zz,Ferrari:1984ozr,Mashhoon:1985cya,Bombelli:1991eg,Cornish:2003ig,Cardoso:2008bp,Dolan:2010wr,Yang:2012he} 
\be \label{wWKB}
\omega^\pm_{\mathrm{WKB}} \approx m\,\Omega_c^{\pm}-i\left(n+\frac{1}{2}\right)\lambda^{\pm}\,,  
\ee
where $n=0,1,2,\ldots$ is the radial overtone.

\begin{figure}
\centering
    \includegraphics[width=0.45\textwidth]{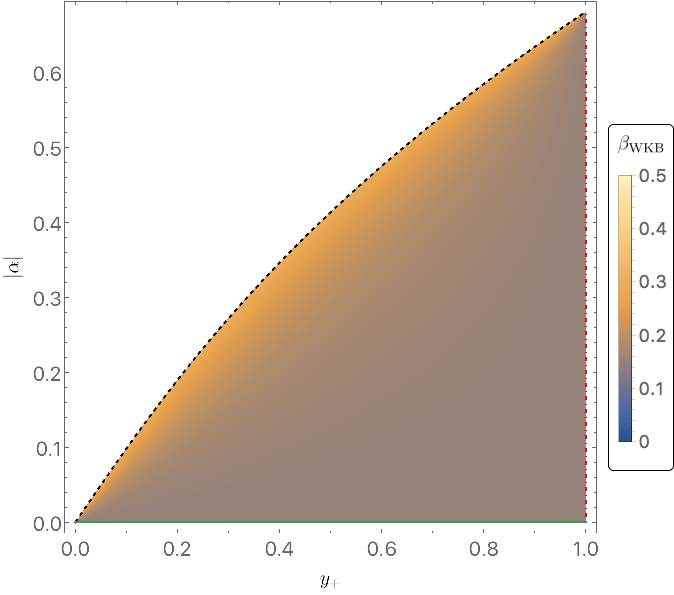}
  \caption{$\beta$ computed in the WKB approximation (using co-rotating photon sphere geodesics) for all values of $(y_+,\alpha)$. $\beta=1/2$ is saturated at extremality, but is otherwise smaller than $1/2$. The extremal curve is represented here by the dashed black line.}
\label{fig:final}
\end{figure}

In Fig.~\ref{fig:final} we plot $\beta_{\mathrm{WKB}}\equiv-{\rm Im}(\omega^+_{\mathrm{WKB}})/\kappa_-$ for $n=0$. For all the range of $(y_+,|\alpha|)$ we find that $\beta_{\mathrm{WKB}}\leq1/2$, with $\beta_{\mathrm{WKB}}=1/2$ saturated only at extremality (represented by the dashed `diagonal' black line in Fig.~\ref{fig:final}). {\it This shows that scalar field perturbations of any non-extremal Kerr-de Sitter black hole respect the strong cosmic censorship conjecture.} 

Of course this calculation was based on approximate (geometric optics/WKB) methods and so one could ask whether corrections to these results might push the true value of $\beta$ above $1/2$, especially near extremality. However, the corrections to ${\rm Im}(\omega)$ are of order $1/|m|$ so, for any fixed background, the corrections can be made arbitrarily small by taking $|m|$ large enough.\footnote{
In fact for vanishing $\Lambda$ the corrections to ${\rm Im}(\omega)$ are ${\cal O}(1/m^2)$ \cite{Dolan:2010wr} and we expect that the same is true with $\Lambda >0$.} So the WKB results should be reliable for sufficiently large $|m|$. In the next section we will determine the quasinormal frequencies numerically and find that, for large enough $|m|$, the WKB result is always in excellent agreement with the exact result. 
%%%%%%%%%%%%%%%%%%%%%%%%%%%%%%%%%%%%%%%%%%%%%%%%%%%%%%%%%%%%
\subsection{\label{sec:scalar}Numerics}
%%%%%%%%%%%%%%%%%%%%%%%%%%%%%%%%%%%%%%%%%%%%%%%%%%%%%%%%%%%%
In this section we will compute numerically the quasinormal modes of a Kerr-de Sitter black hole and make a matching with the analytic results of section \ref{sec:geo}. We first note that the massless scalar wave equation admits separable solutions of the form (\ref{Phi_ans}), with $S_{\omega \ell m}(\chi)$ and $ R_{\omega \ell m}(r)$ obeying the following two-parameter coupled eigenvalue problem
\begin{subequations}\label{eqs:eigen}
\begin{align}
&\partial_{\chi}\left[\Delta_{\chi}(\chi)\partial_{\chi}S_{\omega \ell m}(\chi)\right]-\left[\frac{\Xi^2}{\Delta_{\chi}(\chi)}\left(a\,m-\sigma_\chi\omega\right)^2-K\right]S_{\omega \ell m}(\chi)=0\,,
\\
&\partial_{r}\left[\Delta_{r}(r)\partial_{r}R_{\omega \ell m}(r)\right]+\left[\frac{\Xi^2}{\Delta_{r}(r)}\left(a\,m-\sigma_r\omega\right)^2-K\right]R_{\omega \ell m}(r)=0\,.
\end{align}
\end{subequations}
The symmetry exhibited by the above two equations is only achieved for the particular coordinates used in the line element \eqref{metric}. The eigenvalues to be determined are $(\omega,K)$ where $K$ arises as a separation constant. Before describing the numerical method we used, we first comment on the thorny issue of boundary conditions. Both equations have regular singular points when $\Delta_r$ and $\Delta_\chi$ vanish, so we can use Frobenius method to determine their behaviour there.

For the angular equation, we find
\be
S_{\omega \ell m}(\chi)=(|a|-\chi)^{\pm\frac{|m|}{2}}\sum_{n=0}^{+\infty}(|a|-\chi)^n\,S_{\omega \ell m}^{(n,+)}
\ee
at $\chi=|a|$. Regularity then demands choosing the $+$ sign. A similar behaviour is found at $\chi=-|a|$:
\be
S_{\omega \ell m}(\chi)=(|a|+\chi)^{\pm\frac{|m|}{2}}\sum_{n=0}^{+\infty}(|a|+\chi)^n\,S_{\omega \ell m}^{(n,-)}\,.
\ee
Again the upper sign leads to the physically meaningful solution. We thus conclude that we can factor out all non-analytic behaviour of $S_{\omega \ell m}$ by setting
\be
S_{\omega \ell m}(\chi)=(a^2-\chi^2)^{\frac{|m|}{2}}\tilde{S}_{\omega \ell m}(\chi)\,,
\label{eq:Sre}
\ee
and solving for the smooth eigenfunction $\tilde{S}_{\omega \ell m}(\chi)$.

For the radial coordinate, we have to distinguish the cosmological horizon from the black hole horizon. At the black hole horizon a Frobenius expansion yields
\be
R_{\omega \ell m}(r)=(r-r_+)^{\pm\,\frac{i}{2\,\kappa_+}\left(\omega-m\,\Omega_+\right)}\sum_{n=0}^{+\infty}(r-r_+)^n\,R_{\omega \ell m}^{(n,+)}
\ee
and regularity at the black hole event horizon, which stems from demanding a smooth expansion around $r=r_+$ in ingoing coordinates $(v,r,\chi,\phi')$ at the black hole horizon, demands choosing the lower sign. For the cosmological horizon we find
\be
R_{\omega \ell m}(r)=(r_c-r)^{\pm\,\frac{i}{2\,\kappa_c}\left(\omega-m\,\Omega_c\right)}\sum_{n=0}^{+\infty}(r_c-r)^n\,R_{\omega \ell m}^{(n,c)}\,,
\ee
and again imposing outgoing boundary conditions at the cosmological horizon demands selecting the minus sign in the expression above. We thus consider the following field redefinition:
\be
R_{\omega \ell m}(r)=(r_c-r)^{-\,\frac{i}{2\,\kappa_c}\left(\omega-m\,\Omega_c\right)}(r-r_+)^{-\,\frac{i}{2\,\kappa_+}\left(\omega-m\,\Omega_+\right)}\tilde{R}_{\omega \ell m}(r)
\label{eq:Rre}
\ee
where $\tilde{R}_{\omega \ell m}(r)$ should now be a smooth function with a regular Taylor series at each of the horizons.

The procedure is now clear, we take the field redefinitions (\ref{eq:Sre}) and (\ref{eq:Rre}) and input them into Eqs.~(\ref{eqs:eigen}). The resulting equations are still quadratic in $\omega$ and $K$, and form a coupled eigenvalue problem with eigenfunctions $\big(\tilde{S}_{\omega \ell m}(\chi),\tilde{R}_{\omega \ell m}(r)\big)$ and eigenvalues $(\omega,K)$. The boundary conditions for $\tilde{S}_{\omega \ell m}(\chi)$ and $\tilde{R}_{\omega \ell m}(r)$ are then found by Taylor expanding the equations of motion close to either boundary, and turn out to be of the Robin type, \emph{i.e.}
\be
\mathcal{F}^{1,\pm}(\omega,K)\tilde{S}_{\omega \ell m}^\prime(\pm|a|)=\mathcal{F}^{0,\pm}(\omega,K)\tilde{S}_{\omega \ell m}(\pm|a|)
\ee
and
\be
\mathcal{Q}^{1,+}(\omega,K)\tilde{R}_{\omega \ell m}^\prime(r_+)=\mathcal{Q}^{+,0}(\omega,K)\tilde{R}_{\omega \ell m}(r_+)\,,\quad\mathcal{Q}^{c,1}(\omega,K)\tilde{R}_{\omega \ell m}^\prime(r_c)=\mathcal{Q}^{c,0}(\omega,K)\tilde{R}_{\omega \ell m}(r_c)\,.
\ee
with $\mathcal{F}^{1,\pm}(\omega,K)$, $\mathcal{F}^{0,\pm}(\omega,K)$, $\mathcal{Q}^{1,+}(\omega,K)$, $\mathcal{Q}^{0,+}(\omega,K)$ $\mathcal{Q}^{1,-}(\omega,K)$ and $\mathcal{Q}^{-,+}(\omega,K)$ being known functions which are at most second order polynomials in $\omega$ and $K$. For the numerical procedure, it is also useful to consider coordinates whose range do not depend on the parameters of the solution. To achieve this, we make the following simple linear coordinate transformations
\be
x=\frac{|a|+\chi}{2\,|a|}\qquad\text{and}\qquad y=\frac{1-\frac{r_+}{r}}{1-\frac{r_+}{r_c}}\,.
\ee
The resulting equations are then solved using a Newton-Raphson algorithm, on a unit length Chebyshev grid, as first proposed in \cite{Cardoso:2013pza} and recently detailed in \cite{Dias:2015nua}.

\begin{figure}
\centering
    \includegraphics[width=0.95\textwidth]{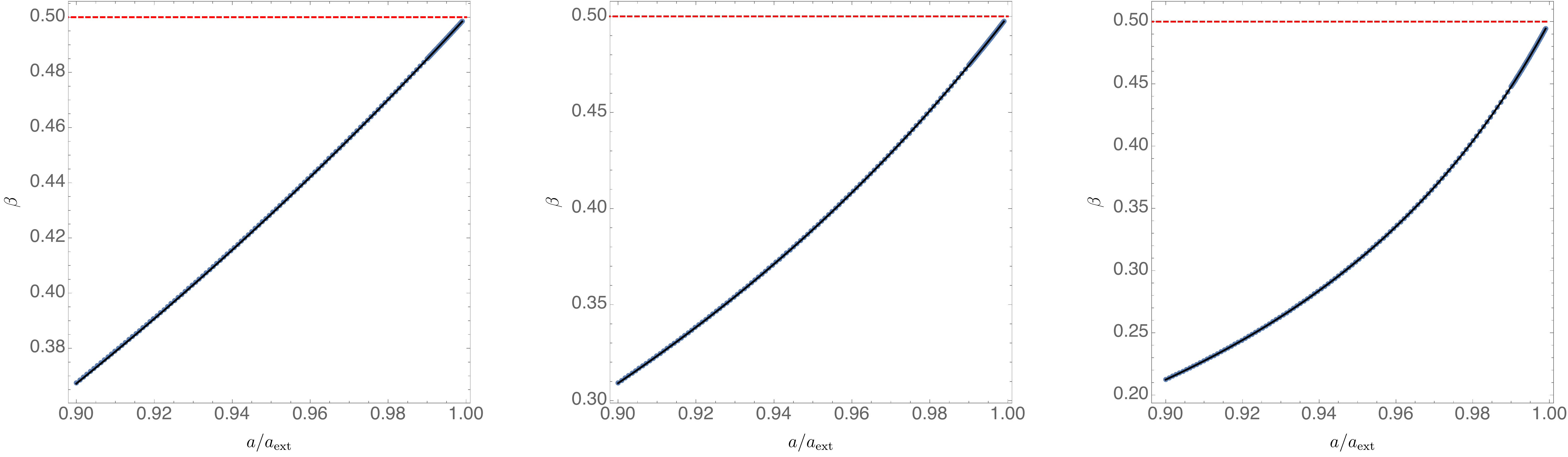}
  \caption{$\beta$ as a function of $a/a_{\mathrm{ext}}$ plotted for fixed several values $y_+=1/4,1/2,3/4$ (from the left to the right panel) and fixed $\ell=m=10$.}
\label{fig:results}
\end{figure}

Our results are shown in Fig.~\ref{fig:results} where we take $y_+=1/4,1/2,3/4$ (from left to right) and plot $\beta$ as a function of $a/a_{\mathrm{ext}}$. Since we are interested in tracking photon sphere modes, we will take $m=\ell=10$. For most of the moduli space of solutions $\beta\ll1/2$, and $\beta$ only gets close to $1/2$ near extremality. This is why in Fig.~\ref{fig:results} we restricted the range of the horizontal axis to $a/a_{\mathrm{ext}}\in[9/10,999/1000]$. Also showing in Fig.~\ref{fig:results} are the analytic WKB photon sphere predictions of section \ref{sec:geo}, see \eqref{wWKB}, denoted by the solid black lines. For $m=\ell=50$ (not shown in Fig.~\ref{fig:results}) we see a maximum deviation between the analytic and numerical data which is not larger than $10^{-6}$ anywhere in parameter space.
To sum up, our numerical results corroborate the analytic analysis performed in section \ref{sec:geo}. For the specific value of $y_+=1/2$, we have pushed our numerical scheme to $1-a/a_{\mathrm{ext}}=10^{-5}$ and see no deviation from the WKB result. 

The absolute error in $\beta$ is not terribly important: what is important is to show that corrections to the WKB result cannot push $\beta$ above $1/2$. To this end we define
\begin{equation}
\Delta \beta \equiv \frac{\beta-\beta_{\mathrm{WKB}}}{\frac{1}{2}-\beta}\,.
\end{equation}
\begin{figure}
\centering
    \includegraphics[width=0.45\textwidth]{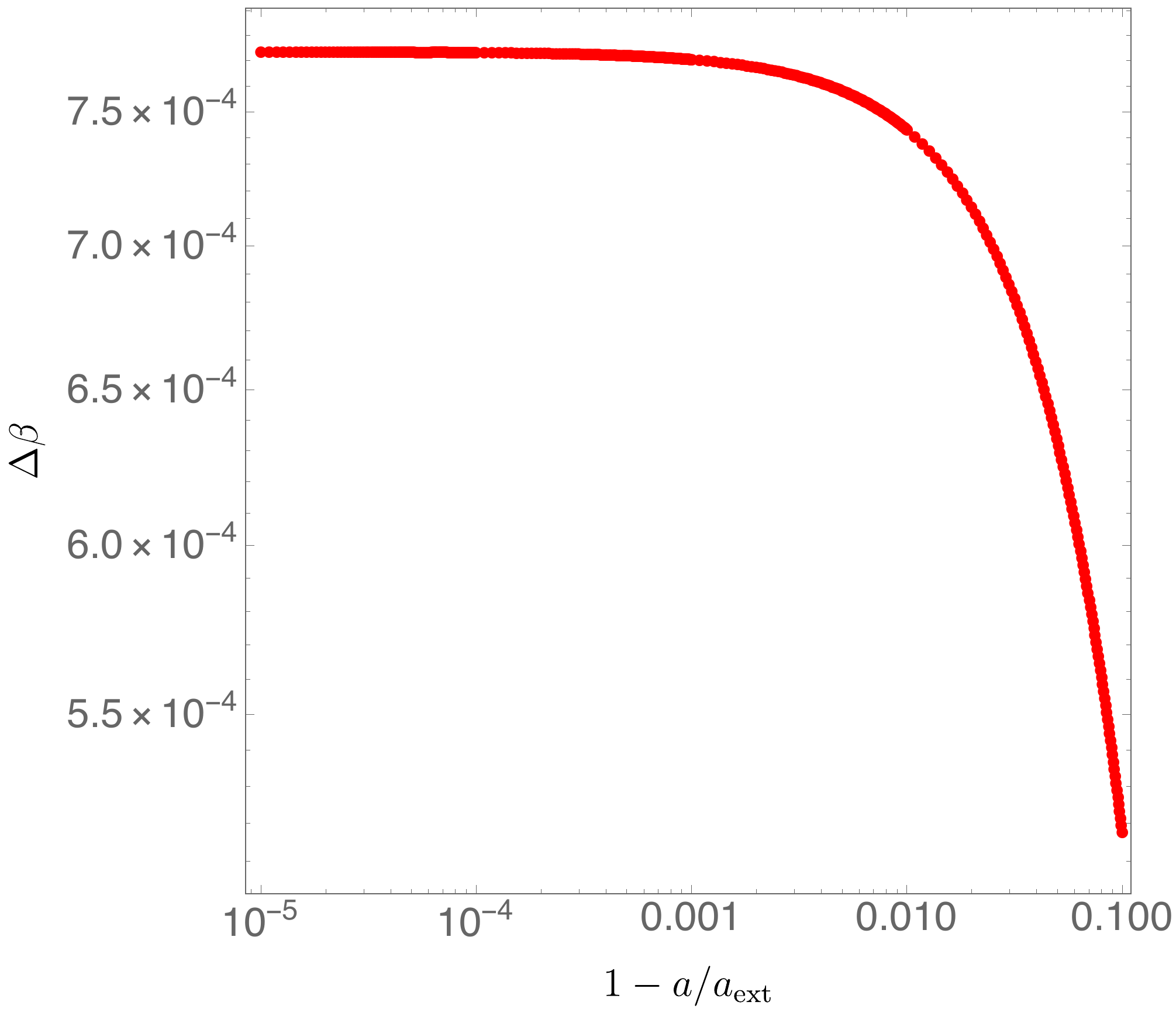}
  \caption{$\log-\log$ plot of $\Delta \beta$ as a function of $1-a/a_{\mathrm{ext}}$ plotted for fixed $y_+=1/2$ and fixed $\ell=m=10$.}
\label{fig:accuracy}
\end{figure}
This quantity is plotted in Fig.~\ref{fig:accuracy} for $y_+ = 1/2$ and $\ell=m=10$. It can be seen that, for $1-a/a_{\rm ext}$ down to $10^{-5}$ we have
\be
 |\Delta \beta | < 10^{-3}.
\ee
Thus the WKB analysis is reliable even very close to extremality. Taking $\ell=m$ to be even larger would make $\Delta \beta$ even smaller.

\section{Gravitational quasinormal modes}

\subsection{Teukolsky equation}

The Kerr-de Sitter black hole is a Petrov type D solution. Therefore, gravitational perturbations of this geometry can be studied using the Teukolsky equation, which uses the Newman-Penrose (NP) framework \cite{Newman:1961qr,Teukolsky:1972my,Teukolsky:1973ha,ChandrasekharBook}. We will study perturbations using the Chambers-Moss null tetrad \eqref{KerrdS:NPtetrad}. For quasinormal modes we assume a separable {\it Ansatz} for the (gauge invariant) perturbed Weyl scalars
\begin{subequations}
\begin{align}
&\psi_0\equiv \NPl^{\mu} \NPm^{\nu}\NPl^{\rho} \NPm^{\alpha} \delta C_{\mu\nu\rho\alpha}=e^{-i\omega t+im \phi}\frac{R^{(+2)}_{\omega \ell m}(r)S^{(+2)}_{\omega \ell m}(\chi)}{(r-i\chi)^2}\,,\label{WeylScalar}
\\
&\psi_4\equiv \NPn^\mu \bar{\NPm}^\nu \NPn^\rho \bar{\NPm}^\alpha \delta C_{\mu\nu\rho\alpha}=e^{-i\omega t+im \phi}\frac{R^{(-2)}_{\omega \ell m}(r)S^{(-2)}_{\omega \ell m}(\chi)}{(r-i\chi)^2}\,,\label{WeylScalar4}
\end{align}
\end{subequations}
where $\delta C_{\mu\nu\rho\alpha}$ are the components of the Weyl tensor perturbation. The Teukolsky equation then reduces to the following two sets of two-parameter eigenvalue problems\footnote{The reader interested on a complete but concise overview that discusses how the solutions of  \eqref{eq:2p}-\eqref{eq:2m} allow to get information about other variables can see section 2 and appendix A of \cite{Dias:2013sdc} (with the trading $L^2\to -L^2$).} 
\begin{equation}
\left\{
\begin{array}{ll}
\displaystyle\left[\mathcal{D}_{-1}\Delta _r\mathcal{D}^{\dagger }_{1 }-6\left(\frac{r^2}{L^2}-i\,\Xi\, \omega\,  r \right)-K^{(+2)} \right]R_{\omega \ell m}^{(+2)}(r)=0\,, &
\\
\\
\displaystyle\left[\mathcal{L}_{-1}\Delta _{\chi }\mathcal{L}^{\dagger }_1-6\left(\frac{\chi ^2}{L^2}+\Xi\,\omega\,\chi \right)+K^{(+2)} \right]S_{\omega \ell m}^{(+2)}(\chi )=0\,, & 
\end{array}
\right.
\label{eq:2p}
\end{equation}
and
\begin{equation}
\left\{
\begin{array}{ll}
\displaystyle\left[\mathcal{D}^{\dagger }_{-1}\Delta _r\mathcal{D}_{1 }-6\left(\frac{r^2}{L^2}+i\,\Xi\, \omega\,  r \right)-K^{(-2)} \right]R_{\omega \ell m}^{(-2)}(r)=0\,, &
\\
\\
\displaystyle\left[\mathcal{L}^{\dagger }_{-1}\Delta _{\chi }\mathcal{L}_1-6\left(\frac{\chi ^2}{L^2}-\Xi\,\omega\,\chi \right)+K^{(-2)} \right]S_{\omega \ell m}^{(-2)}(\chi )=0\,, & 
\end{array}
\right.
\label{eq:2m}
\end{equation}
where $K^{(\pm2)}$ are separation constants and we defined the operators \cite{ChandrasekharBook,Chambers:1994ap}:
\begin{eqnarray}
 && \hspace{-1.5cm}  \mathcal{D}_n=\partial _r+i\,\frac{\Xi}{\Delta _r} \left(m a-\omega \sigma_r\right)+n\,\frac{\partial_r\Delta _r}{\Delta _r},\qquad \mathcal{D}^{\dagger }_n=\partial_r-i\,\frac{\Xi}{\Delta_r}\left(m a-\omega \sigma_r\right)+n\,\frac{\partial_r \Delta _r}{\Delta _r} \,,\nonumber\\
 && \hspace{-1.5cm} \mathcal{L}_n=\partial _{\chi }+\frac{\Xi}{\Delta _{\chi }} \left(m a -\omega \sigma_\chi\right)+n\,\frac{\partial_\chi\Delta _{\chi }}{\Delta _{\chi }}, \qquad \mathcal{L}^{\dagger}_n=\partial _{\chi }-\frac{\Xi}{\Delta _{\chi }} \left(m a -\omega \sigma_\chi\right)+n\,\frac{\partial_\chi\Delta _{\chi }}{\Delta _{\chi }}\,.
 \label{diffOp}
\end{eqnarray}
Equations (\ref{eq:2p}) and (\ref{eq:2m}) are isospectral\footnote{We have explicitly checked this is the case, by computing the corresponding sets of quasinormal modes associated with each of the equations.}, that is to say, once appropriate boundary conditions are imposed, they give the same values of $\omega$ and $K^{(+2)}=K^{(-2)}$. So in the following section, we shall focus on the pair $\{R^{(+2)}_{\omega \ell m}(r),S^{(+2)}_{\omega \ell m}(\chi)\}$ with eigenvalues $\{\omega,K^{(+2)}\}$.

For our discussion of strong cosmic censorship, we need to determine the behaviour of  the Weyl scalar $\psi_0$ defined in  \eqref{WeylScalar} at the Cauchy horizon. For that we use the outgoing coordinates $(u,r,\chi,\phi'')$ that extend the solution across $r=r_-$. The radial equation for $R^{(2)}_{\omega \ell m}(r)$ has a regular singular point when  $\Delta_r=0$ and thus a Frobenius analysis yields the two possible behaviours at the Cauchy horizon $r=r_-$. We find that the most general solution for $\psi_0$ near $r=r_-$ is a linear combination of $\psi_0^{(1)}$ and $\psi_0^{(2)}$ where
\begin{subequations}
\begin{align}
& \psi_0^{(1)} = e^{-i \omega u} e^{i m \phi''} S^{(+2)}_{\omega \ell m}(\chi) (r-i \chi )^{-2} (r-r_-) \hat{R}^{(+2)(1)}_{\omega \ell m}(r)\,, \\
& \psi_0^{(2)} = e^{-i \omega u} e^{i m \phi''} S^{(+2)}_{\omega \ell m}(\chi) (r-i \chi )^{-2} (r-r_-)^{ -1+i (\omega - m \Omega_-)/\kappa_-} \hat{R}^{(+2)(2)}_{\omega \ell m}(r)\,,
\end{align}
\end{subequations}
where $\Omega_- = \Omega(r_-)$ and $\hat{R}^{(+2)(1)}_{\omega \ell m}$, $\hat{R}^{(+2)(2)}_{\omega \ell m}$ are smooth functions of $r$ that are non-zero at $r=r_-$.

This gives the behaviour in the Chambers-Moss tetrad \eqref{KerrdS:NPtetradout}. This tetrad is not regular at the Cauchy horizon so we need to convert our results to a regular tetrad. Consider the Weyl scalar $\widetilde{\psi}_0\equiv \widetilde{\NPl}^{\mu} \widetilde{\NPm}^{\nu}\widetilde{\NPl}^{\rho} \widetilde{\NPm}^{\alpha} \delta C_{\mu\nu\rho\alpha}$ defined using the regular null tetrad $\{ \widetilde{\NPl},\widetilde{\NPn},\widetilde{\NPm},\widetilde{\NPbm}\}$ defined in \eqref{KerrdS:NPtetradoutSmooth}. We now have $\widetilde{\psi}_0 = \psi_0/\Delta_r$ and hence, near the Cauchy horizon $\widetilde{\psi}_0$ is a linear combination of $\widetilde{\psi}_0^{(1)}$ and $\widetilde{\psi}_0^{(2)}$, where
\begin{subequations}
\begin{align}
& \widetilde{\psi}_0^{(1)} = e^{-i \omega u} e^{i m \phi''} S^{(+2)}_{\omega \ell m}(\chi) (r-i \chi )^{-2}  \tilde{R}^{(+2)(1)}_{\omega \ell m}(r)\,, \\
& \widetilde{\psi}_0^{(2)} = e^{-i \omega u} e^{i m \phi''} S^{(+2)}_{\omega \ell m}(\chi) (r-i \chi )^{-2} (r-r_-)^{ -2+i (\omega - m \Omega_-)/\kappa_-} \tilde{R}^{(+2)(2)}_{\omega \ell m}(r)\,,
\end{align}
\end{subequations}
and $\tilde{R}^{(+2)(i)}_{\omega \ell m} \equiv F \hat{R}^{(+2)(i)}_{\omega \ell m}$ $(i=1,2)$ where $F \equiv (r-r_-)/\Delta_r$ is smooth and non-vanishing at the Cauchy horizon. It follows that the $\tilde{R}^{(+2)(i)}_{\omega \ell m}$ are smooth and non-vanishing at the Cauchy horizon. 

The solution $\widetilde{\psi}_0^{(1)}$ is smooth and non-vanishing at the Cauchy horizon. However, the solution $\widetilde{\psi}_0^{(2)}$ diverges at the Cauchy horizon. A quasinormal mode solution will be a linear combination of these two solutions and there is no reason why either coefficient in this linear combination should vanish. It follows that $\widetilde{\psi}_0$ diverges at the Cauchy horizon. Defining $p = i (\omega - m \Omega_-)/\kappa_-$, the behaviour, in the regular tetrad, of a quasinormal mode near the Cauchy horizon is
\be \label{psi0key}
\widetilde{\psi}_0\sim (r-r_-)^{p -2}.
 \ee
We now define $\beta={\rm Re}(p)=-{\rm Im}(\omega)/\kappa_-$ as before. We will show that if the quasinormal mode corresponds to a linearized metric perturbation that, in some gauge, is in $H^1_{\rm loc}$ then we must have $\beta \ge 1/2$. 

The easiest way to see this is as follows. If the linearized metric perturbation is in $H^1_{\rm loc}$ in some gauge then its second derivative belongs to $H^{-1}_{\rm loc}$ \cite{folland}. Hence it must be possible to interpret the (gauge invariant) quantity $\widetilde{\psi}_0$ as a tempered distribution in $H^{-1}_{\rm loc}$. The latter is the dual space of $H^1_{\rm loc}$ \cite{folland} so if $\widetilde{\psi}_0$ belongs to $H^{-1}_{\rm loc}$ then $\int \widetilde{\psi}_0 f$ should be finite for any $f \in H^1_{\rm loc}$. Choose compactly supported $f$ where the support of $f$ contains a segment of the Cauchy horizon with $f$ smooth except on this segment, with $f \sim (r-r_-)^q$ in a neighbourhood of this segment, where $q$ is real. This $f$ belongs to $H^1_{\rm loc}$ if, and only if, $q>1/2$. Then $\int \widetilde{\psi}_0 f$ converges for all $q>1/2$ if, and only if, $\beta \ge 1/2$. Hence if (\ref{psi0key}) belongs to $H^{-1}_{\rm loc}$ then we must have $\beta \ge 1/2$. 

A less rigorous, argument goes as follows. Assume that, in the coordinates $(u,r,\chi,\phi'')$ each component of the metric perturbation behaves (near the Cauchy horizon) as $(r-r_-)^q$ where $q= q_R+i\,\big({\rm Re}(\omega)-m\Omega_-\big)/\kappa_-=q_R + i \,{\rm Im}(p)$. The value of the real part $q_R$ may be different for different components. The condition that the perturbation belongs to $H^1_{\rm loc}$ is that each component must have $q_R > 1/2$. Since the Weyl tensor perturbation involves two derivatives of the metric, it follows that the Weyl scalar must be at least as smooth as $(r-r_-)^{{\rm min}(q_R)+i\,{\rm Im}(p)-2}$ and hence from (\ref{psi0key}) we must have ${\rm Re}(p) >{\rm min}(q_R)>1/2$, \emph{i.e.}, $\beta > 1/2$.

We conclude that if a quasinormal mode corresponds to a linearized metric perturbation that, in some gauge, belongs to $H^1_{\rm loc}$ then the mode must have
\be
 \beta  \ge  \frac{1}{2} \qquad {\rm where} \qquad \beta \equiv -\frac{{\rm Im}(\omega)}{\kappa_-}.
\ee 
Hence if {\it all} gravitational quasinormal modes have $\beta \ge  1/2$ then strong cosmic censorship might be violated. However, if we can find one quasinormal mode with $\beta <1/2$ then, as argued in the Introduction, a generic linearized gravitational perturbation cannot be extended across the Cauchy horizon in $H^1_{\rm loc}$ and so strong cosmic censorship holds.

We can use geometric optics to calculate the frequencies of ``photon sphere" gravitational quasinormal modes with $\ell = |m| \gg 1$. The calculation is exactly as in section \ref{sec:geo}. As explained in \cite{Dolan:2010wr} (see Eq.~(51) of \cite{Dolan:2010wr}), the spin dependence of the WKB approximation of quasinormal frequencies with $\ell=|m|$ only comes at order $1/m$. This makes sense, since in the WKB limit we are taking $\ell=|m|$ to be large, while keeping the spin fixed (either to zero, in the scalar case, or to two in the gravitational case). Hence for $\ell = |m| \gg 1$, the gravitational quasinormal frequencies are, to leading order, exactly the same as the scalar field quasinormal frequences, as computed in section \ref{sec:geo}. Furthermore, the subleading terms in $\beta$ can be made arbitrarily small by taking $\ell=|m|$  large enough.

We conclude that for any non-extremal Kerr-de Sitter black hole, there are gravitational quasinormal modes with $\beta < 1/2$. Hence {\it linearized gravitational perturbations of any non-extremal Kerr-de Sitter black hole respect the strong cosmic censorship conjecture}. 

In the next section, we will check the accuracy of the geometric optics/WKB approximation for gravitational perturbations by computing the quasinormal frequencies numerically. Just as for the scalar field case, we will find that the geometric optics approximation is always very accurate for $\ell=m \gg 1$. 

%%%%%%%%%%%%%%%%%%%%%%%%%%%%%%%%%%%%%%%%%%%%%%%%
%%%%%%%%%%%%%%%%
\subsection{Numerics}

We write the perturbation for the Weyl scalar $\psi_0$ as in \eqref{WeylScalar}. Our task now is to find $S^{(+2)}_{\omega \ell m}(\chi)$, $R^{(+2)}_{\omega \ell m}(r)$ and the eigenvalues $\{\omega,K^{(+2)}\}$ by solving \eqref{eq:2p}. As explained before,  \eqref{eq:2m} is isospectral to  \eqref{eq:2p} and thus we do not consider it further.

This section follows \emph{mutatis mutandis} section \ref{sec:scalar}, so we will only point out the differences.   
Regularity at the poles, located at $\chi = \pm|a|$ now demands that
\be
S^{(+2)}_{\omega \ell m}(\chi) = (a^2-\chi^2)^{\frac{|m-2|}{2}}\tilde{S}^{(+2)}_{\omega \ell m}(\chi)\,,
\ee
where $\tilde{S}^{(+2)}_{\omega \ell m}(\chi)$ is a smooth function of $\chi$ for all values of $m$. We have discarded the irregular solution $ (a^2-\chi^2)^{-\frac{|m-2|}{2}}$. Demanding outgoing boundary conditions at the cosmological horizon $-$ \emph{i.e.} that the solution is regular at $r=r_c$ in outgoing coordinates  $(u,r,\chi,\phi'')$  $-$  and ingoing boundary conditions at the black hole horizon  $-$ \emph{i.e.} that the solution is regular at $r=r_+$ in ingoing coordinates $(v,r,\chi,\phi')$  $-$  now motivates the following field redefinition:
\be
R^{(+2)}_{\omega \ell m}(r)=(r_c-r)^{-\,\frac{i}{2\,\kappa_c}\left(\omega-m\,\Omega_c\right)+1}(r-r_+)^{-\,\frac{i}{2\,\kappa_+}\left(\omega-m\,\Omega_+\right)-1}\tilde{R}^{(+2)}_{\omega \ell m}(r)\,,
\ee
where again $\tilde{R}^{(+2)}_{\omega \ell m}(r)$ is a smooth function at both $r=r_+$ and $r=r_c$.  

\begin{figure}[h]
\centering
    \includegraphics[width=0.45\textwidth]{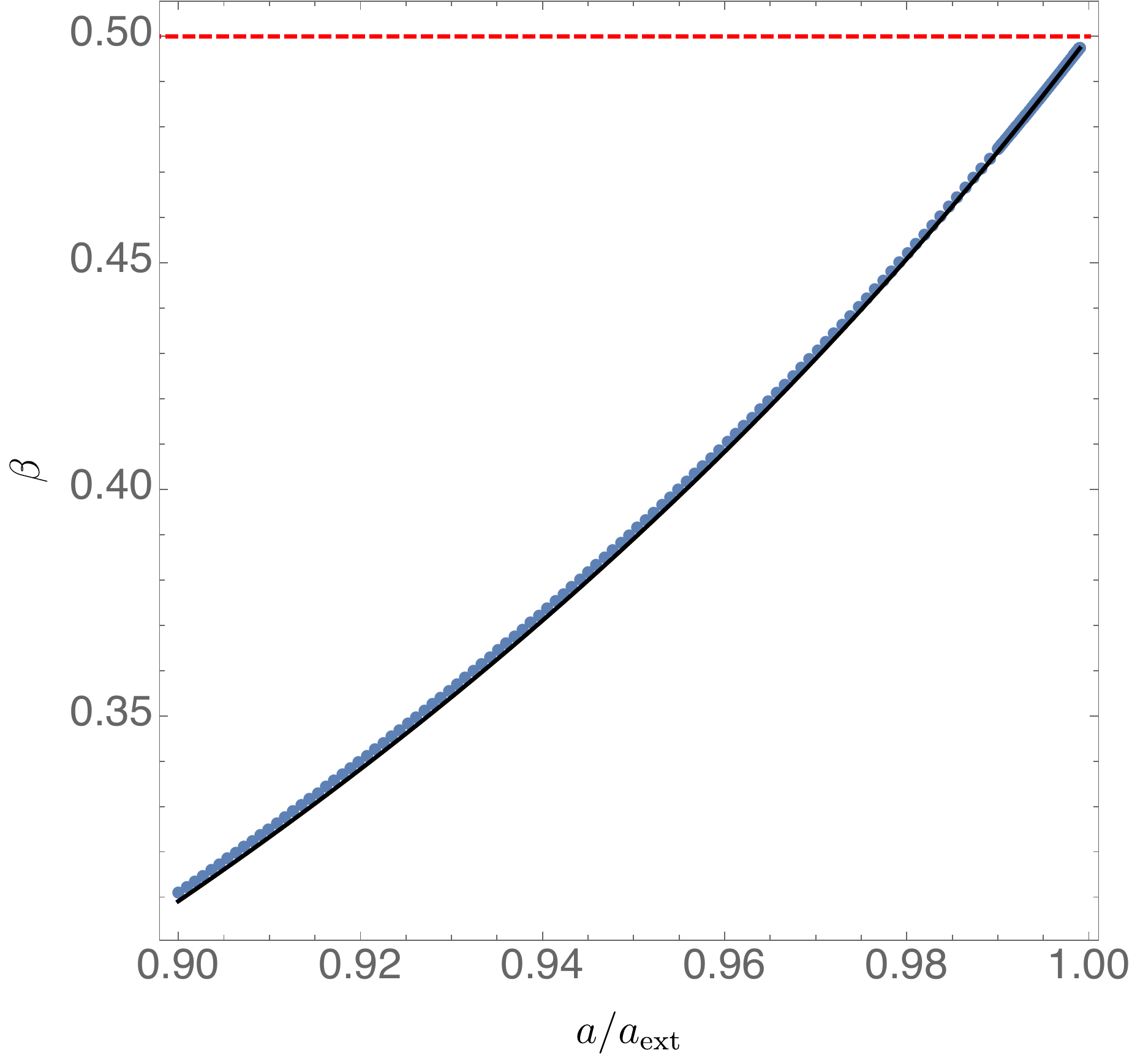}
  \caption{$\beta$ for gravitational perturbations as a function of $a/a_{\mathrm{ext}}$ plotted for fixed $y_+=1/2$ and $\ell=m=10$.}
\label{fig:quasigrav}
\end{figure}

The numerical results are displayed in Fig.~\ref{fig:quasigrav}, where we have choosen $y_+=1/2$, $a/a_{\mathrm{ext}}\in[9/10,999/1000]$ and $m=\ell=10$. As expected, at large enough $\ell=m$, the spin is irrelevant, and the analytic approximation of section \ref{sec:geo} is excellent. The only difference worth noticing is that it seems we need to get to larger values of $m=\ell$ in order for the geometric optics approximation to be as accurate as for the scalar field case. We note, however, that the approximation remains reliable even as we approach extremality.
%%%%%%%%%%%%%%%%%%%%%%%%%%%%%%%%%%%%%%%%%%%%%%%%
%%%%%%%%%%%%%%%%

\end{document}